\documentclass[universe,article,accept,moreauthors,pdftex]{mdpi}
\firstpage{1}
\pubvolume{7}
\issuenum{12}
\articlenumber{462}
\pubyear{2021}
\copyrightyear{2020}
\externaleditor{Academic Editor: Norma G. Sanchez} 
\datereceived{16 October 2021}
\dateaccepted{23 November 2021}
\datepublished{27 November 2021}
\hreflink{https://doi.org/10.3390/universe 
7120462}

 \def \kms {\ifmmode  \,\rm km\,s^{-1} \else $\,\rm km\,s^{-1}  $ \fi }
\def \kpc {\ifmmode  {\rm kpc}  \else ${\rm  kpc}$ \fi  }

\def\apj{Astrophys. J.}%
\def\apjl{Astrophys. J. Lett.}%
\def\prd{Phys.~Rev.~D}%
\def\prl{Phys.~Rev.~Lett.}%
\def \sci {Science}
\Title{Warm Dark Matter from Higher-Dimensional Gauge Theories}
\TitleCitation{Warm Dark Matter from Higher-Dimensional Gauge Theories}

\Author {Sinziana Paduroiu $^{1}$*\orcidA{}, Michael Rios $^{2}$\orcidB{}, Alessio Marrani $^{3}$\orcidC{} and David Chester $^{4}$}
\AuthorNames{Sinziana Paduroiu, Michael Rios, Alessio Marrani and David Chester}
\AuthorCitation{Paduroiu, S.; Rios, M.; Marrani, A.; Chester, D.}
\address{$^{1}$ \quad QOOSP Lab, Las Vegas, 89169, USA\\
$^{2}$ \quad Dyonica ICMQG, Los Angeles, 90032, USA; michael@dyonica.org\\
$^{3}$ \quad Centro Studi e Ricerche Enrico Fermi, Rome, 00184, Italy; alessio@dyonica.org\\
$^{4}$ \quad Quantum Gravity Research, Los Angeles, 90290, USA; davidc@quantumgravityresearch.org}

\corres{Correspondence: sinziana@qoosplab.org}

\abstract{Warm dark matter particles with masses in the keV range have been linked with the large group representations in gauge theories through a high number of species at decoupling. In this paper, we address WDM fermionic degrees of freedom from such representations. Bridging higher-dimensional particle physics theories with cosmology studies and astrophysical observations, our approach is two-folded, i.e., it includes realistic models from higher-dimensional representations and constraints from simulations tested against observations. Starting with superalgebras in exceptional periodicity theories, we discuss several symmetry reductions and we consider several representations that accommodate a high number of degrees of freedom. We isolate a model that naturally accommodates both the standard model representation and the fermionic dark matter in agreement with both large and small-scale constraints. This model considers an intersection of branes in $D=27+3$ in a manner that provides the degrees of freedom for the standard model on one hand and 2048 fermionic degrees of freedom for dark matter, corresponding to a $\sim$2 keV particle mass, on the other. In this context, we discuss the theoretical implications and the observable predictions.}

\keyword{warm dark matter; fermions; degrees of freedom; higher-dimensional gauge theories; M-theory}

\begin{document}
\section{Introduction}

In the past decade, having been thoroughly tested against the data from observations, the cold dark matter (CDM) model
has encountered difficulties in explaining these data and has proved to be inconsistent with the observations under several aspects. Besides the two initial concerns---the large cores in galaxies and the small number of satellites---that may or may not be reconciled by introducing finely tuned baryonic processes in the simulations, there are several other points of concern at both small and large scales, e.g., the fact that the population of dwarf galaxies in voids is in contradiction with the predictions from CDM simulations~\cite{tikhonov09, zavala09, peebles2010}; the failure to replicate, in CDM simulations, the high number of pure bulgeless galaxies observed and the low number of galaxy mergers~\cite{Kormendy10, Kormendy15}; the formation and growth of supermassive black holes in agreement with observed distributions~\cite{PLB}; and the recent detection of high-redshift quasars~\cite{Wang}.

WDM scenarios with particles in the keV range are gaining ground in explaining some of these discrepancies and many studies show the advantages that a keV dark matter model presents in explaining the observations at both small and large scales~\cite{HdV10, HdV12, HdV12b, HdV12c, Destri13, Destri13b, Destri13c, Paduroiu1, Paduroiu, HdV14, Sanchez16}. A keV WDM particle, for example, can explain the existence of cores in galaxies of varied masses, from dwarf galaxies to spirals, when considering the quantum Fermi pressure of the particles~\cite{Destri13, Destri13b, Destri13c}. Moreover, while, for spiral galaxies, in an attempt to reconcile the CDM model with observations, the presence of the cores has been attributed to the stars and gas concentration at the center of galaxies---tentatively described by adding baryonic components in the simulations, with different studies showing different results, sometimes contradictory (e.g.,~\cite{mac, Marinacci})---for dwarf galaxies, which are dark matter dominated, this cannot be the case.

Furthermore, in WDM models with keV particles, stars can form early on inside the filaments, as shown in unparalleled high-resolution simulations~\cite{Gao} and could provide the seeds for the early formation of black holes, predictions that could be tested with more such high-resolution simulation studies.

Given the challenges that the CDM is facing and in the absence of a successful experimental detection of a CDM particle candidate, studies of warm dark matter models are highly justified both from the theoretical and numerical perspectives.

From the point of view of particle physics, the theoretical models that predict candidates for the generically called `warm dark matter' are more varied than those for CDM, with different production mechanisms that translate into several intrinsic properties of the dark matter particles, reviewed in \cite{boy19}. The consensus to what these particles have in common is that they possess an astronomically significant free streaming length. The free streaming length implies a cutoff in the power spectrum and a non-zero velocity dispersion. The value for these `quantities' is dictated by the particular model and production mechanism and it gives a certain mass for the considered particle. In~\cite{Paduroiu1, Paduroiu}, several WDM particle models and their assumptions have been re-examined, showing how strongly model-dependent their properties are and, consequently, how weak the constraints on the particle mass from the over-simplified simulations are.

Therefore, the catch comes from the freedom WDM provides in terms of the particle candidates that are theorized in this category, mainly in the mass-velocity correspondence. Since, in reality, we can only constrain, with observations and simulations, the velocity dispersion of the particle and not the mass, other particle physics constraints have to be considered in the absence of detection experimental results.

A particular model that has been extensively used in warm dark matter simulations is the one described in~\cite{bode01}, which implies a much higher number of degrees of freedom at decoupling than for the standard model and its supersymmetric extensions~\cite{Pierpaoli}. As hinted in~\cite{bode01}, this would be possible in larger gauge groups, especially those with extra dimensions.

Several studies have looked at larger gauge group representations in relation to dark matter~\cite{Jedamzik, Gross, so16xso16, DixonHarvey, McGuigan}, but none that would explain the high number of degrees of freedom. In the context of the so-called {\it exceptional} super Yang Mills theories, large group representations that could allow this high number to be reached have been explored~\cite{geoESYM, ESYMtheory}. In this paper, we revise these results and attempt to develop and isolate models that would produce the required high number of degrees of freedom at decoupling, while preserving the standard model representations. We found that these representations favor 2048 fermionic degrees of freedom at decoupling, which corresponds to a $\sim$2 keV particle, which happens to be in agreement with observations.

In Section~\ref{2}, we review the warm dark matter model requirement of such high degrees of freedom. In Section~\ref{3}, we introduce the algebras with higher dimensional spinors and discuss several reductions in the context of exceptional periodicity. In Section~\ref{4}, we discuss the WDM and SM disentanglement through braneworld intersections, isolating a mechanism through which SM degrees of freedom are isolated from WDM ones and WDM becomes neutral under the photon gauging the
electromagnetic $U(1)$. Our conclusions are summarized in Section~\ref{5}.

%%%%%%%%%%%%%%%%%%%%%%%%%%%%%%%%%%%%%%%%%%
\section{keV WDM from Higher-Dimensional Gauge Theories \label{2}}

Within the WDM scenario, several models with various particle production mechanisms and various properties have been previously studied. We can distinguish between thermally produced neutrinos and those coming from a non-resonant production mechanism~\cite{Dolgov, Dodelson, Shi, Asaka, boy09} for particles in the eV--hundreds keV range that do not decay. More complex models that consider decay~\cite{Fuller} and re-thermalization~\cite{Lello} have also been considered as well as quantum semi-degenerate particles with a non-negligible chemical potential~\cite{Paduroiu1}.

However, in numerical simulations, the varied models are oversimplified, often using a ‘cutoff’ in the power spectrum as the only parameter that distinguishes WDM from CDM particles, as reviewed in~\cite{Paduroiu21}. The model most favored by simulation studies for a thermal relic WDM particle, which provides this ‘cutoff’ mass-velocity, is the one described in~\cite{bode01}. Below, we discuss this model, its assumptions and its implications.

\subsection*{WDM Degrees of Freedom}

The particle assumed in~\cite{bode01} is a thermal relic particle that
decouples while still relativistic, but, by equality, is nonrelativistic. The
streaming speed at equality $v_{s}/c \propto T_{X}/m_{X}$, where the
effective temperature $T_{X}$ varies inversely with the scale factor. In
order to have a streaming speed that is low enough to keep up with the CDM
success at large scales---meaning, the formation of structures at an early
enough redshift and within the observable mass scales for halos---the
decrease in the effective temperature is suggested~\cite{bode01}. If the
particle decouples with thermal abundance, $\rho_{X} \propto T_{X}^3m_{X}$
must be held fixed. Therefore, to obtain a reasonable reduction in
velocity, one needs to decrease the temperature and increase the mass. One
way the reduction in temperature can be realized is through increasing the
number of the degrees of freedom at decoupling. The example given in \cite%
{bode01} assumes a reduction in $T_{X}$ of $\sim$4 and an increase in mass
of $4^3$, thus corresponding to 688 degrees of freedom. However, for 1 keV,
the number of degrees of freedom is higher and this is usually what
is indirectly used in simulations when the velocity of the WDM particle is
introduced, as shown in~\cite{Paduroiu1, Paduroiu}.

Below, we remind the derivation in~\cite{bode01} and the discussion in \cite%
{Paduroiu1}. As recalled in~\cite{bode01} (see \cite{bode01} Appendix A for a detailed derivation and \mbox{\cite{Paduroiu1}} for a detailed assessment of that derivation), for a
thermal relict particle X that decouples when relativistic, the abundance $%
n_X$ relative to photons is
\begin{equation}  \label{eq:abund}
\frac{n_X}{n_\gamma}=\left(\frac{43/4}{g_\mathrm{dec}}\right)\left(\frac{4}{%
11}\right)\frac{g_X}{2}\ ,
\end{equation}

\noindent where $g_\mathrm{dec}$ is the number of relativistic species
present at decoupling---degrees of freedom---and $g_X$ is the number of spin
states of the particle. Assuming that the distribution function scales as
the non-thermal distribution $(\exp(v/v_0)+1)^{-1}$, for a redshift $z$, the
velocity $v_0$ (in Equation (A3)~\cite{bode01}) is given,
\begin{equation}  \label{eq:velocities}
\frac{v_0(z)}{1+z} = \mathbf{0.012}\left(\frac{\Omega_{X}}{0.3}\right)^{%
\frac{1}{3}}\left(\frac{h}{0.65}\right)^{\frac{2}{3}} \left(\frac{1.5}{g_{X}}%
\right)^{\frac{1}{3}}\left(\frac{\mathrm{keV}}{m_{X}}\right)^{\frac{4}{3}}%
\kms \ .
\end{equation}

The relation between the cosmological parameters and the particle's density $%
\rho_{x}$ is
\begin{equation}  \label{eq:omega}
\Omega_{X}h^2\equiv\frac{\rho_x}{\rho_c} h^2 =\frac{m_Xn_X}{3H^2/8\pi G}
h^2\approx\frac{115}{g_\mathrm{dec}}\frac{g_{X}}{1.5}\frac{m_{X}}{\mathrm{keV%
}}\ ,
\end{equation}
for the Hubble constant $H \equiv 100 h \,\mathrm{km \,s}^{-1} \,\mathrm{kpc}%
^{-1}$.

Therefore, for an $m_X=1$\,keV particle, considering $\Omega_{X}=0.3$ as in \cite%
{bode01}, we have the following formula for the velocity:
\begin{equation}  \label{eq:final}
\frac{v_0(z)}{1+z}\approx{\mathbf{0.12}\left(\frac{1}{g_\mathrm{dec}}%
\right)^{1/3}\frac{\mathrm{keV}}{m_{X}}}\kms \
\end{equation}

\noindent that gives $g_{\mathrm{dec}}\approx \mathbf{1000}%
\,(g_{X}/1.5)^{1/3}$ for a 1 keV particle.

This large value for the number of degrees of freedom at decoupling is much
higher than the value for the standard model, $\sim$107, and even the
supersymmetric extension, $\sim$229~\cite{Pierpaoli}. Other studies
arbitrarily choose, in their analyses, a much more conservative value for the
number of degrees of freedom of the WDM particle candidates, i.e., $\sim$150
for a massive neutrino~\cite{Pierpaoli}; $\sim$100 for right-handed
neutrinos that would decouple before the electro-weak phase transition \cite%
{Colombi}. The number of degrees of freedom scales with the mass of the
particle, therefore a higher mass would correspond to a higher number of
degrees of freedom (and a stronger decrease in temperature). Since this
value is correlated to the mass--velocity correspondence that dictates the
behavior of the WDM particle and its influence on the structure formation
and evolution, we study the theoretical models where such a large number of
degrees of freedom can be achieved.

While the authors in~\cite{bode01} do hint at larger group representations
with extra dimensions in relation to the high $g_\mathrm{dec}$ number, they
also mention that entropy-producing processes that take place after the
decoupling of the particle can  mimic an increase in $g_\mathrm{dec}
$. Indeed, trying to avoid such large group representations,~\cite{boy19,
Bezrukov, Nemevsek, Patwardhan} discuss the entropy generation by the
out-of-equilibrium decay of heavier particles, which would allow the
number of degrees of freedom of the keV particle to be similar to that used
for active neutrinos~\cite{Bezrukov}. Self-interactions as a possible way to
reduce the number densities by self-annihilations are also discussed in \cite%
{Herms}. However, in this paper, we look at the large gauge group
representations that would naturally produce this high number of degrees of
freedom, starting with the symmetry reductions in algebras with
higher-dimensional spinors in exceptional periodicity theories.

%%%%%%%%%%%%%%%%%%%%%%%%%%%%%%%%%%%%%%%%%%
\section{Algebras with Higher-Dimensional Spinors \label{3}}

A class of algebras involving higher-dimensional spinors is
studied in the so-called \textit{exceptional }super Yang--Mills (ESYM)
theories, firstly discussed in~\cite{geoESYM,ESYMtheory}, in
which higher-dimensional theories \cite%
{VafaF,BarsS,Bars14,Sezgin,Nishino,1bis} beyond string/M-theory \cite%
{WittenM,matrixM} have been considered and extended, also in relation to
exceptional periodicity (EP) algebras \cite%
{amsMS,group32p,group32a}. In this section, we provide various
branchings of these algebras with a number of spinor
degrees of freedom of $\mathcal{O}(10^{3})$.
Here, we discuss the symmetry reductions along the worldvolume (WV) of p-branes, which centrally extend
the underlying chiral superalgebra and the reductions that follow the so-called \textit{Magic Star} structure---named so in~\cite{amsMS} after the star polygon geometric shape---of the corresponding EP algebra.

\subsection{Superalgebra in $17+1$ and 9-Brane WV reduction}

One of the simplest examples is based on the $(1,0)$ superalgebra obtained
by setting $\mathbf{n}=1$ in Equation (3.5) of~\cite{geoESYM}, namely, the chiral
superalgebra of ESYM in $D=s+t=17+1$ space--time dimensions, whose central
extensions are given by the r.h.s.~of the  anticommutator

\begin{equation}
\left\{ Q_{\alpha },Q_{\beta }\right\} =\left( \gamma ^{\mu }\right)
_{\alpha \beta }P_{\mu }+\left( \gamma ^{\mu _{1}...\mu _{5}}\right)
_{\alpha \beta }Z_{\mu _{1}...\mu _{5}}+\left( \gamma ^{\mu _{1}...\mu
_{9}}\right) _{\alpha \beta }Z_{\mu _{1}...\mu _{9}},  \label{alg1}
\end{equation}

\noindent namely, by a 1-brane (fundamental string), an electric 5-brane and
its dual, a magnetic 9-brane. The corresponding EP 3-graded algebra, whose
semisimple, 0-graded part yields the homogeneous Lorentz algebra of $17+1$
space--time, is obtained by setting $n=1$ in (II.14) of \cite{group32a}%
, namely,
\begin{equation}
\mathfrak{e}_{6(-26)}^{(1)}:=\mathbf{256}_{-1}^{\prime }\oplus \left(
\mathfrak{so}_{17,1}\oplus \mathbb{R}\right) _{0}\oplus \mathbf{256}_{1},
\label{EP1}
\end{equation}

\noindent where $\mathbf{256}$ and $\mathbf{256}^{\prime }$, respectively,
denote the Majorana--Weyl (MW) semispinor in $17+1$ dimensions and its
conjugate (note that the EP level $n$ is understood with $+1$ in~\cite{geoESYM}, namely,
the first non-trivial level of EP is given by $n=2$ therein). It should be remarked that, when the spinors are interpreted as
Abelian translational generators, the non-negatively graded part of the EP
algebra $\mathfrak{e}_{6(-26)}^{(1)}$~(\ref{EP1}) yields the Lie algebra of
the global isometry group of the homogeneous non-symmetric real special
manifold $L(16,1,0)$ in the classification cited in~\cite{dWVP, dWVVP}.

Symmetry reduction along the WV of the highest-dimensional central extension
occurring in the r.h.s.~of (\ref{alg1}), namely, along the $9+1$ dimensional
WV of the 9-brane, yields the breaking
\begin{eqnarray}
\mathfrak{so}_{17,1} &\rightarrow &\mathfrak{so}_{9,1}\oplus \mathfrak{so}%
_{8}\oplus (\mathbf{10},\mathbf{8}_{v});  \label{br1} \\
\mathbf{256} &=&\left( \mathbf{16},\mathbf{8}_{s}\right) \oplus \left(
\mathbf{16}^{\prime },\mathbf{8}_{c}\right) ;  \label{br2} \\
\mathbf{256}^{\prime } &=&\left( \mathbf{16}^{\prime },\mathbf{8}_{s}\right)
\oplus \left( \mathbf{16},\mathbf{8}_{c}\right) ,  \label{br3}
\end{eqnarray}

\noindent thus yielding an $\mathfrak{so}_{8}$ inner/fiber symmetry, whose
only  peculiar polarization is shown in the branchings (\ref{br1})--(%
\ref{br3}).

Let us recall that, in the models for family unification beyond the SM, such as the ones
discussed in~\cite{Wilczek,BTZee}, one posits a further breaking of the
inner symmetry,
\begin{eqnarray}
\mathfrak{so}_{8} &\rightarrow &\mathfrak{so}_{5}\oplus \mathfrak{so}%
_{3}\oplus (\mathbf{5},\mathbf{3})\simeq \mathfrak{usp}_{4}\oplus \mathfrak{%
su}_{2}\oplus (\mathbf{5},\mathbf{3});  \label{br4} \\
\mathbf{8}_{v} &=&\left( \mathbf{5},\mathbf{1}\right) \oplus \left( \mathbf{1%
},\mathbf{3}\right) ;~\mathbf{8}_{s}=\left( \mathbf{4},\mathbf{2}\right) ;~~%
\mathbf{8}_{c}=\left( \overline{\mathbf{4}},\mathbf{2}\right) ,  \label{br6}
\end{eqnarray}

\noindent where $\mathfrak{so}_{3}$ is used for family symmetry \cite%
{Wilczek}.

\subsection{Superalgebra in $20+4$ and 12-Brane WV Reduction}

Another example is provided by the \textit{quasi-conformal} level of the
above symmetry, namely, by the the $(1,0)$ superalgebra obtained by setting $%
\mathbf{n}=1$ in Equation (3.8) of~\cite{geoESYM}, the chiral superalgebra of
ESYM in $D=20+4$ space--time dimensions, whose central extensions are given
by the r.h.s.~of the  anticommutator
\begin{equation}
\left\{ Q_{\alpha },Q_{\beta }\right\} =\eta _{\alpha \beta }Z+\left( \gamma
^{\mu _{1}...\mu _{4}}\right) _{\alpha \beta }Z_{\mu _{1}...\mu _{4}}+\left(
\gamma ^{\mu _{1}...\mu _{8}}\right) _{\alpha \beta }Z_{\mu _{1}...\mu
_{8}}+\left( \gamma ^{\mu _{1}...\mu _{12}}\right) _{\alpha \beta }Z_{\mu
_{1}...\mu _{12}},  \label{alg2}
\end{equation}

\noindent namely, by a 0-brane, a 4-brane, an electric 8-brane and its dual,
a magnetic 12-brane. The corresponding 5-graded EP algebra is obtained by
setting $n=1$ in (II.16) of~\cite{group32a}
\begin{eqnarray}
\mathfrak{e}_{8(-24)}^{(1)}:= &&\mathfrak{so}_{20,4}\oplus \mathbf{2048}
\label{EP2} \\
=&&\mathbf{22}_{-2}\oplus \mathbf{1024}_{-1}^{\prime }\oplus \left(
\mathfrak{so}_{19,3}\oplus \mathbb{R}\right) _{0}\oplus \mathbf{1024}%
_{1},\oplus \mathbf{22}_{2},  \label{EP3}
\end{eqnarray}

\noindent where $\mathbf{2048}$ and $\mathbf{1024}$ 
denote the MW
semispinors in $20+4$ and $19+3$ dimensions, respectively. Again, when the
spinors and vectors are interpreted as Abelian translational generators, the
non-negatively graded part of the EP algebra $\mathfrak{e}_{8(-24)}^{(1)}$ (%
\ref{EP3}) corresponds to the Lie algebra of the global isometry group of
the homogeneous non-symmetric quaternionic K\"{a}hler manifold given by the
image of the aforementioned space $L(16,1,0)$ under the composite c$\cdot $%
R-map~\cite{dWVP, dWVVP}.

The symmetry reduction along the WV of the highest-dimensional central extension
occurring in the r.h.s.~of (\ref{alg2}), namely the $12+4$ dimensional WV of
the 12-brane, yields the~breaking\vspace{-6pt} 
\begin{eqnarray}
\mathfrak{so}_{20,4} &\rightarrow &\mathfrak{so}_{12,4}\oplus \mathfrak{so}%
_{8}\oplus (\mathbf{16},\mathbf{8}_{v});  \label{Br1} \\
\mathbf{2048} &=&\left( \mathbf{128},\mathbf{8}_{s}\right) \oplus \left(
\mathbf{128}^{\prime },\mathbf{8}_{c}\right) ,  \label{Br2}
\end{eqnarray}

\noindent thus again yielding an $\mathfrak{so}_{8}$ inner/fiber symmetry, where $%
\mathbf{128}$ and $\mathbf{128}^{\prime }$ stand for the MW semispinor in $%
12+4$ dimensions and its conjugate, respectively.

\subsubsection{Further Reduction in Inner/Fiber Symmetry}

Through a disentangling mechanism discussed in Section~\ref{4}, one copy of
the $\mathbf{128}$ encodes SM matter and the rest provides thousands of
fermionic degrees of freedom.

A possible way to do this is to further break the inner/fiber symmetry $%
\mathfrak{so}_{8}$ down to $\mathfrak{so}_{7}$ and/or $\mathfrak{g}_{2}$:\vspace{-6pt} 
\begin{eqnarray}
\mathfrak{so}_{20,4} &\rightarrow &\mathfrak{so}_{12,4}\oplus \mathfrak{so}%
_{8}\oplus (\mathbf{16},\mathbf{8}_{v})\rightarrow \mathfrak{so}%
_{12,4}\oplus \mathfrak{so}_{7}\oplus (\mathbf{1},\mathbf{7})\oplus (\mathbf{%
16},\mathbf{7})\oplus (\mathbf{16},\mathbf{1}); \\
\mathbf{2048} &=&\left( \mathbf{128},\mathbf{8}_{s}\right) \oplus \left(
\mathbf{128}^{\prime },\mathbf{8}_{c}\right) =\left( \mathbf{128},\mathbf{8}%
\right) \oplus \left( \mathbf{128}^{\prime },\mathbf{8}\right) ,
\end{eqnarray}
\noindent or
\begin{eqnarray}
\mathfrak{so}_{20,4} &\rightarrow &\mathfrak{so}_{12,4}\oplus \mathfrak{so}%
_{8}\oplus (\mathbf{16},\mathbf{8}_{v})\rightarrow \mathfrak{so}%
_{12,4}\oplus \mathfrak{so}_{7}\oplus (\mathbf{1},\mathbf{7})\oplus (\mathbf{%
16},\mathbf{7})\oplus (\mathbf{16},\mathbf{1})  \nonumber \\
&\rightarrow &\mathfrak{so}_{12,4}\oplus \mathfrak{g}_{2}\oplus 2\cdot (%
\mathbf{1},\mathbf{7})\oplus (\mathbf{16},\mathbf{7})\oplus (\mathbf{16},%
\mathbf{1}); \\
\mathbf{2048} &=&\left( \mathbf{128},\mathbf{8}_{s}\right) \oplus \left(
\mathbf{128}^{\prime },\mathbf{8}_{c}\right) =\left( \mathbf{128},\mathbf{8}%
\right) \oplus \left( \mathbf{128}^{\prime },\mathbf{8}\right)   \nonumber \\
&=&\left( \mathbf{128},\mathbf{7}\right) \oplus \left( \mathbf{128}^{\prime
},\mathbf{7}\right) \oplus \left( \mathbf{128},\mathbf{1}\right) \oplus
\left( \mathbf{128}^{\prime },\mathbf{1}\right) .
\end{eqnarray}

While this isolates a $\mathbf{128}$ as a singlet of $\mathfrak{g}_{2}$ and
has additional fermions, breaking $\mathfrak{so}_{12,4}$ further to identify
the standard model and electromagnetism would lead to the additional
fermions to have the same charge,  therefore not be dark matter.
Nevertheless, the utilization of the $\mathbf{2048}$ spinor in $\mathfrak{e}%
_{8(-24)}^{(1)}$ is later used to identify a candidate for WDM.

\subsubsection{\textit{Magic Star} Reduction in $D=20+4$}

\begin{figure}[H]
\includegraphics[scale=0.8]{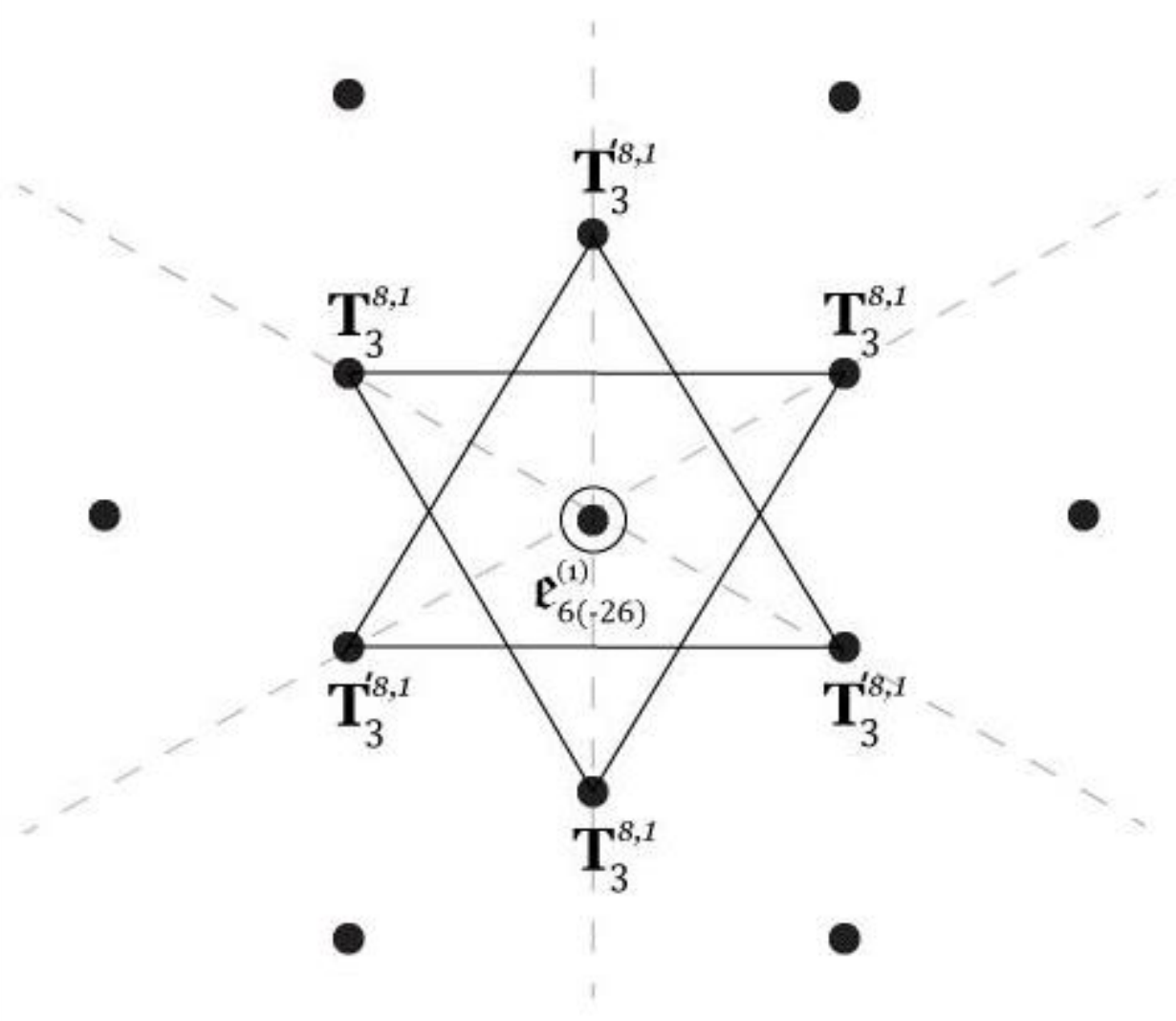}
\caption{A \textit{Magic Star}-type projection of  $\mathfrak{e}_{8(-24)}^{(1)}$ that contains $\mathfrak{e}_{6(-26)}^{(1)}$ in the center, named after the star polygon geometric shape.}
\label{fig:MagicStarT_8-1-e6}
\end{figure}

Another possible symmetry reduction on the same ESYM, which is useful in
order to discuss family unification beyond the SM, concerns the exploitation
of the so-called \textit{Magic Star}~\cite{amsMS} decomposition of the EP
algebra $\mathfrak{e}_{8(-24)}^{(1)}$ (\ref{EP2})-(\ref{EP3}), with the EP
algebra $\mathfrak{e}_{6(-26)}^{(1)}$ (\ref{EP1}) posited in the center of
the star-like arrangement of (generalized) roots,

\begin{equation}
\mathfrak{e}_{8(-24)}^{(1)}=\mathfrak{e}_{6(-26)}^{(1)}\oplus \mathfrak{sl}%
_{3}(\mathbb{R})\oplus \left( \mathbf{T}_{3}^{8,1},\mathbf{3}\right) \oplus
\left( \left( \mathbf{T}_{3}^{8,1}\right) ^{\prime },\mathbf{3}^{\prime
}\right) ,
\end{equation}

\noindent as depicted in Figure~\ref{fig:MagicStarT_8-1-e6}, where $\mathbf{T}%
_{3}^{8,1}$ denotes the Hermitian part of the special Vinberg cubic
T-algebra~\cite{Vinberg,Alek} with elements realized as formal $3\times 3$
Hermitian matrices, whose manifestly $\mathfrak{so}_{16}$-covariant
structure reads

\begin{equation}
X:=\left(
\begin{array}{ccc}
\mathbf{1}_{I} & \mathbf{16} & \mathbf{128} \\
\ast  & \mathbf{1}_{II} & \mathbf{128}^{\prime } \\
\ast  & \ast  & \mathbf{1}_{III}%
\end{array}%
\right) \in \mathbf{T}_{3}^{8,1},  \label{T}
\end{equation}

\noindent where $\mathbf{128}$ and $\mathbf{128}^{\prime }$, respectively,
denote the MW semispinor in $16$ dimensions and its conjugate (the case $q=8$
and $n=1$ of Equation (III.1) of~\cite{group32a}). In addition, it is here worth
recalling that the Hermitian matrix-like structure (\ref{T}) corresponds to
the manifestly $\mathfrak{so}_{17,1}$-covariant Peirce decomposition

\begin{equation}
\mathbf{T}_{3}^{8,1}=\mathbf{1}_{-4}\oplus \mathbf{18}_{2}\oplus \mathbf{256}%
_{-1},
\end{equation}

\noindent where the $\mathfrak{so}_{17,1}$-singlet corresponds to the $%
\mathfrak{so}_{16}$-singlet $\mathbf{1}_{III}$ in (\ref{T}).

\section{\textls[-15]{Symmetry Reduction to the Standard Model and Warm Dark Matter~Disentanglement} \label{4}}

By studying $E_8\times E_8$ heterotic string theory on the boundary of an
11-dimensional supergravity bulk, two $E_8$ gauge theories on the endpoints
of an orbifold from $\mathbb{R}^{10}\times S^1/\mathbb{Z}_2$ have been found
\cite{HoravaWitten1,HoravaWitten2}. The anomaly cancellation requires a
twelve-form, which is factorized into a 4-form and 8-form, yet these are
invoked in 10- and 11-dimensional theories, implying a complete description
from larger dimensions. Heterotic string theory must have an origin from $%
D=25+1$ or higher. While bosonic string theory exists in \mbox{$D=25+1$}, heterotic
string theory places 16 dimensions on an even, self-dual lattice and includes fermions. $%
D=26+1$ bosonic M-theory and $D=26+2$ target space--time for 2D gravity has
been introduced~\cite{HorowitzSusskind,Watabiki}. How $D=11+3$ unifies the
superalgebras of type I, type IIA, type IIB and heterotic theories has been
articulated in~\cite{BarsS, Bars14, Sezgin}. The superalgebras of M-theory
and F-theory can arise from nested braneworlds in $D=27+3$, as the
exceptional Yang--Mills superalgebra contains an 11-brane worldvolume with
spinors~\cite{ESYMtheory}. Dimensional reduction also finds a 10-brane
worldvolume with a 16D transverse space and includes~spinors.

The Horava--Witten theory~\cite{HoravaWitten1,HoravaWitten2} shows how M-theory
on $\mathbb{R}^9\times S^1 \times S^1/\mathbb{Z}_2$, when compactified on $S^1
$, leads to type IIA and $E_8\times E_8$ heterotic string theories, which are
T-dual to type I and $Spin(16)\times Spin(16)\subset Spin(32)$ superstring
theories. This implies that the type I superstring on $\mathbb{R}^9\times S^1
$ naturally has a $Spin(16)\times Spin(16)$ vacuum. Dark matter has been
discussed in relation to nonsupersymmetric $Spin(16)\times Spin(16)$ string
theory~\cite{so16xso16,DixonHarvey,McGuigan}. While heterotic string theory
leading to the supersymmetric standard model allows for supersymmetric light
dark matter candidates such as neutralinos 
\cite{Jungman}, light
supersymmetric partners of the standard model below 1.5 TeV have not been
observed~\cite{ATLAS}. The exceptional Yang--Mills superalgebras facilitate an understanding of various worldvolume constructions for M-theory and beyond,
which allows for supersymmetric forms of matter beyond the standard model
that are different from the superpartners or the nonsupersymmetric dark
matter models.

Since the membranes discussed in this section are more complicated than those contained within M-theory itself, a brief pedagogical introduction to membranes and their relationship to gauge theory is warranted. 
To start, let us consider the worldline formulation of a free point particle described by the following action:
\begin{equation}
S_0 = -mc\int ds = -mc\int \sqrt{g_{\mu\nu} \dot{x}^\mu \dot{x}^\nu} d\tau.
\end{equation}

\noindent Varying this action leads to the standard geodesic equation in general relativity. 
Extended objects in supergravity lead to ``p-brane'' for arbitrary-dimensional membranes, which spans a $(p+1)$-dimensional 
worldvolume and generalizes the worldline formulation above~\cite{Duff}. 
While strings are 1-branes on a worldsheet and M-theory has a 2-brane worldvolume, 3-brane worldvolumes were found in type IIB supergravity~\cite{Cederwall:1996pv}.
F-theory admits 3-brane phenomenology that can include the standard model with gravity~\cite{HeckmanVafa}. 
Furthermore, 3-branes are  intimately related to $AdS_5$, whose isometry group is the conformal group $Spin(4,2)$. 
Additionally, $AdS_5$ supports a 4D CFT with a 3-brane on the boundary, such that $dS_4$ can be 
obtained within the Randall--Sundrum model~\cite{RandallSundrum,Hirayama:2006jn}.

Before the full significance of the Yang--Mills theory for the standard model was established, 
the nonlinear sigma model was introduced for pions, which is a nonrenormalizable theory similar to general relativity. 
General relativity is equivalent to a nonlinear sigma model Lagrangian~\cite{CheungRemmen}, which is similarly used for the bosonic sector of p-branes:

\end{paracol}
\begin{equation}
S_{p}=\int d^{p+1}x\left[ \frac{1}{2}\sqrt{\left\vert g\right\vert }\left(
g^{\mu \nu }G_{MN}\partial _{\mu }X^{M}\partial _{\nu }X^{N}-1\right) +\frac{%
1}{2\left( p+1\right) !}\epsilon ^{\mu _{1}...\mu
_{p+1}}C_{M_{1}...M_{p+1}}\partial _{\mu _{1}}X^{M_{1}}...\partial _{\mu
_{p+1}}X^{M_{p+1}}\right],
\end{equation}
\begin{paracol}{2}
\switchcolumn

\noindent where $C_{M_1\dots M_{p+1}}$ is a $p+1$-form potential sourced by the p-brane and $X^M$ is a scalar field with an index with respect to the target space, such as $D=10+1$ in M-theory. 
The bosonic p-brane Lagrangian in the static gauge leads to a Born--Infeld action, 

\begin{equation}
S_p = -T\int d^{p+1}x \sqrt{-\mbox{det}\left(\eta_{\mu\nu}+\frac{1}{T}\partial_\mu X^M \partial_\nu X^N G_{MN}\right)},
\end{equation}

\noindent where $T=1/(2\pi l_P)^p$ is the brane tension. The Born--Infeld action was introduced as an 
electrodynamical theory; it is the 1-loop renormalization of QED. The historical developments 
of transitioning from the non-renormalizable nonlinear sigma model to the renormalizable Yang--Mills 
theory suggest that the Yang--Mills gauge gravity theories, such as the MacDowell--Mansouri theory, 
will remain important for the unification of the standard model and gravity~\cite{MacDowellMansouri}. As braneworlds are 
intimately related to the study of gravity and beyond-the-standard-model physics is necessary for dark matter, 
we propose that braneworld spinors allow us to consider new dark matter candidates. 

We briefly review the representation theory of the standard model as well as the most common high-energy GUT models, recalling that $SU(3)_c \times U(1)_e \subset SU(5) \subset Spin(10) \subset E_{6(-78)} \subset E_{8(-24)}$. $E_{6(-78)}$ allows us to consider three charts of $Spin(10)\times U(1)$, such that $E_{8(-24)}$ contains three charts of $Spin(10)\times Spin(2,4)$ for a combination of GUT and conformal symmetries. The $D=12+4$ (1,0) chiral super-Poincar\'e algebra also exists and contains a ${\bf 128}$ spinor. When breaking to $Spin(2,4)\times Spin(10)$, the three conformal charts span $Spin(4,4)$ (associated to a conformal braneworld symmetry), such that three distinct ${\bf16}$ representations of $Spin(10)$ are found, each leading to the standard model fermion representations with respect to $SU(3)_c \times SU(2)_L \times U(1)_Y$,

\begin{equation}
{\bf 16} = (\overline{\mathbf{3}},\mathbf{1})_{\frac{1}{3}}\oplus(\mathbf{1},\mathbf{2})_{-\frac{1}{2}}\oplus(\mathbf{3},\mathbf{2})_{\frac{1}{6}}\oplus(\overline{\mathbf{3}},\mathbf{1})_{-\frac{2}{3}}\oplus(\mathbf{1},\mathbf{1})_1\oplus(\mathbf{1},\mathbf{1})_0.
\end{equation}

Once the bosons are unified into a single gauge group, such as $SU(5)$ or $Spin(10)$, the vector bosons are described by the Yang--Mills action

\begin{equation}
S_{YM} = \int d^dx \sqrt{-g} \frac{1}{g_{YM}^2}F_{\mu\nu}^A F_{\rho\sigma}^A g^{\mu\rho}g^{\nu\sigma}.
\end{equation}

At high energies, an $SU(2,2)\times U(1)$ conformal gauge group, axial $U(1)$ and $SU(5)$ GUT leads to $SU(7,2)$, which is also a maximal subalgebra of $E_{8(-24)}$. An $E_{8(-24)}$ gauge theory does not need to be considered, but rather gauge theories from subalgebras. For instance, $E_6$ GUT contains a complex ${\bf 27}$ and $\overline{{\bf 27}}$ in combination with the adjoint representation ${\bf 78}$~\cite{Gursey}. 

Combined with another $U(1)$, the total of all of these states leads to $E_7$ by combining the adjoint bosonic sector with the complex fermionic sector, 

\begin{equation}
{\bf 133} = {\bf 27} \oplus {\bf 78}\oplus{\bf 1} \oplus \overline{{\bf27}}.
\end{equation}

\noindent As shown above, the representations found within $E_{6}$ GUT stem from a single adjoint representation of $E_{7}$; similarly, $E_8$. Recently, using $E_6$ to obtain the standard model spectrum has been explored~\cite{Boyle} without using $E_6$ GUT. $E_8$ for three generations of matter has also been explored~\cite{BarsGunaydin}, although conformal spacetime symmetry was not considered. This non-supersymmetric $E_8$ GUT model and the supersymmetric $E_8\times E_8$ heterotic string theory use this to make contact with three generations of matter. Since two of the authors recently found EP algebras that generalize $E_8$ ~\cite{amsMS},  we expand the exploration of the representations associated with these algebras. Dark matter is an immediate natural candidate, as the EP algebras contain spinors for matter beyond the standard model. Superalgebras are explored to motivate braneworld intersections that naturally lead to the standard model spectrum as described above, along with additional dark matter candidates. 

\subsection{Braneworld Spinors}

$D=26+1$ M-theory with a $\mathbf{8,192}$ spinor~\cite{MonstrousM} and many p-branes provides
an origin for heterotic string theory and M-theory. A 2-brane, 5-brane,
10-brane, 13-brane, 18-brane and 21-brane is found from considering a superalgebra with $\mathbf{8,192}$ supercharges.  $D=10+1$ M-theory has a 2-brane and 5-brane, while a 21-brane was anticipated in $D=26+1$ bosonic M-theory \cite%
{HorowitzSusskind}.  In $D=26+1$, a 10-brane and 18-brane allow for a
(10,1) worldvolume theory and a (18,1) worldvolume theory
. Breaking $%
Spin(26,1)$ to various subgroups makes contact with worldvolume and transverse
symmetries of interest,

\begin{equation}
\begin{array}{c|c|c|c}
\mbox{Braneworld} & \mbox{Worldvolume} & \mbox{Transverse} & \mbox{Spinors}
\\ \hline
\mbox{2-brane} & Spin(2,1) & Spin(24) & (\mathbf{2},~\mathbf{2048}^{\prime
})\oplus (\mathbf{2},~\mathbf{2048}) \\
\mbox{10-brane} & Spin(10,1) & Spin(16) & (\mathbf{32},\mathbf{128}^{\prime
})\oplus (\mathbf{32}^{\prime },\mathbf{128}) \\
\mbox{18-brane} & Spin(18,1) & Spin(8) & (\mathbf{512},\mathbf{8}^{\prime
})\oplus (\mathbf{512}^{\prime },\mathbf{8})%
\end{array}%
\end{equation}

\noindent in which the breaking of the $\mathbf{8,192}$ spinor is also
provided. The 2-brane relates to the Horowitz--Susskind \cite%
{HorowitzSusskind} $D=2+1$ CFT with $SO(24)$ global symmetry and $AdS_4\times S^{23}$ holography. The spinors in the 10-brane worldvolume only see the $%
\mathbf{32}$ spinor of M-theory from $(\mathbf{32},\mathbf{128})$ components. The $D=26+1$ theory contains a $%
\mathbf{128}$ spinor of a transverse $Spin(16)$, just as the superstring has
an $\mathbf{8}_{s}$ spinor of $Spin(8)$. This can be compared to the Horava--Witten theory~\cite{HoravaWitten1, HoravaWitten2} with $\mathbb{R}%
^{10}\times S^{1}/\mathbb{Z}_{2}$, by orbifolding a spatial direction of the 10-brane.
The 18-brane gives a $D=18+1$ ``M-theory" that reduces to a $D=17+1$ string
theory as studied in relation to Kac--Moody algebras~\cite{West}.
Quasiconformal extensions of the 2-brane, 10-brane, 18-brane and global
space--time motivate $Spin(4,4)$, $Spin(12,4)$, $Spin(20,4)$ and $Spin(28,4)$ symmetries and their EP extensions.

$Spin(12,4)$ is the conformal symmetry of $Spin(11,3)$, whose exceptional
Yang--Mills superalgebra admits a Hodge duality between a 3-brane and 7-brane when the 3-brane is treated as a
worldvolume for space--time, generalizing the 2-brane and 5-brane. The
isometry group of the 3-brane (3,3) worldvolume has $Spin(4,4)$ conformal symmetry, which
describes three generations of matter~\cite{Kostant,CMR}. A transverse $Spin(8)$
is used for visible matter within $Spin(12,4)$, where fermionic degrees of freedom can be encoded by a $\mathbf{128}$ spinor of
$Spin(12,4)$, leading to $\mathfrak{e}_{8(-24)}$. Going to $Spin(20,4)$ or $Spin(28,4)$ allows us to consider transverse $Spin(16)$ or $Spin(24)$ for possible dark matter

$D=28+4$ entails the following braneworld twistors for three generations:
\begin{equation}
\begin{array}{c|c|c|c}
\mbox{Braneworld} & \mbox{Conformal worldvolume} & \mbox{Transverse} & %
\mbox{Spinors} \\ \hline
\mbox{3-brane} & Spin(4,4) & Spin(24) & (\mathbf{8},~\mathbf{2048}^{\prime
})\oplus (\mathbf{8}^{\prime },~\mathbf{2048}) \\
\mbox{11-brane} & Spin(12,4) & Spin(16) & (\mathbf{128},\mathbf{128}^{\prime
})\oplus (\mathbf{128}^{\prime },\mathbf{128}) \\
\mbox{19-brane} & Spin(20,4) & Spin(8) & (\mathbf{2048},~\mathbf{8}^{\prime
})\oplus (\mathbf{2048}^{\prime },~\mathbf{8})%
\end{array}%
\end{equation}

Taking the 11-brane and 19-brane in $D=27+3$, they are forced to intersect spatially via a 3-brane. The 11-brane isolates a $D=11+3$ worldvolume with a 3-brane and dual 7-brane, while the 19-brane isolates a $D=19+3$ worldvolume with a 3-brane and dual
15-brane. Our proposal is that the $D=11+3$ worldvolume theory encodes baryonic matter, while the $D=19+3$ theory encodes a dark matter sector.

EP algebras provide new dark matter models via intersected or nested 
 higher
membrane worldvolumes.  In order to combine dark matter and the standard
model within the same space--time worldvolume, $\mathfrak{e}_{8(-24)}^{(2)}$ with $\mathfrak{so}_{28,4}$ is of
interest, as it contains $\mathfrak{so}_{12,4}\oplus\mathfrak{so}_{16}$ 
in addition to various vector and spinor
representations. The group $Spin(12,4)$ combines the 4D conformal group with $Spin(10)$ GUT. 
The compact real form of $\mathfrak{e}_8^{(2)}$ also contains $\mathfrak{so}_{16}\oplus%
\mathfrak{so}_{16}$, which has been studied in relation to dark matter in a different context~\cite{McGuigan}.

\subsection{Braneworld Intersections for Warm Dark Matter beyond the Standard Model}

One candidate for describing the standard model from string theory is 
the utilization of D3 and D7 branes from type IIB supergravity or F-theory
\cite{D3D7}. The central extension of $D=11+3$ minimal chiral $(1,0)$
superalgebra naturally contains a 3-brane and dual 7-brane, as well as the type IIB superalgebra \cite%
{Bars14,Sezgin}. The role of $E_{8(-24)}$ with $Spin(12,4)$ in the context of
background-independent two-time M-theory and heterotic string theory with $%
\mathcal{N}=2$ or $1$ has been mentioned in~\cite{Martinec}. A single
generation of fermions can be found via $\mathbf{64}$ of $Spin(11,3)$ as
off-shell degrees of freedom~\cite{Krasnov}. While breaking $%
Spin(12,4)\rightarrow Spin(2,4)\times Spin(10)$ is tempting, this makes it
difficult to identify the three generations of matter with less than $%
\mathbf{2048}$ spinors without another consideration. Some of the authors
have explored how $SU(3,2)\times SU(5)$ (maximal and non-symmetric subgroup
of $E_{8(-24)}$) naturally leads to three conformal charts, in a manner
similar to how quasiconformal groups with four times are the conformal group
for three generations of matter~\cite{CMR}. The corresponding (1,0) chiral super-Poincar\'e algebra in $D=12+4$ entails three
generations of the standard model within $\mathbf{128}$ of $Spin(12,4)$
instead of 192 off-shell degrees of freedom.

Throughout this section, it is assumed that three generations of the
standard model stem from a 3-brane and dual 7-brane, immersed in an
11-brane worldvolume. To go beyond the standard model and include dark
matter candidates, a 15-brane is also considered. The 3-brane and 15-brane exist inside a 19-brane worldvolume. The possible $p$-branes in $D=27+3$ are
shown in Figure~\ref{brane-intersections}. In particular, visible matter and
dark matter are isolated on the 11-brane and 19-brane, such that they
intersect on the $11+19-27=3$-brane worldvolume for space--time. In this manner,
the braneworld intersection identifies visible and dark matter with
different forces, yet they exist within the same space--time. The $D=27+3$ theory may be isolated to $D=19+3$ for dark matter, yet reduced
to $D=11+3$ for visible~matter.

Here, a model via $Spin(12,4)\times Spin(16)$ is pursued.  The $Spin(16)$ factor arises from the transverse symmetry of an 11-brane worldvolume in $D=27+3$.  Then, we study a braneworld intersection
of the 11-brane and 19-brane in $D=27+3$.

\textls[-15]{By restricting to the 11-brane worldvolume, the 3-brane is dual to a
7-brane. Isolating the conformal group of the 11-brane worldvolume within $%
\mathfrak{e}_{8(-24)}^{(2)}$ encodes a transverse $Spin(16)$,}
%\newpage
\end{paracol}
\begin{eqnarray}
\underset{\mathbf{496\oplus 32,768}}{\mathfrak{e}_{8(-24)}^{(2)}}
&\rightarrow &\mathfrak{so}_{28,4}\rightarrow \mathfrak{so}_{12,4}\oplus
\mathfrak{so}_{16}\rightarrow \mathfrak{so}_{4,4}\oplus \mathfrak{so}%
_{8}\oplus \mathfrak{so}_{16}, \\
\mathbf{496} &=&(\mathbf{120},\mathbf{1})\oplus (\mathbf{1},\mathbf{120}%
)\oplus (\mathbf{16},\mathbf{16})  \nonumber \\
&=&(\mathbf{28},\mathbf{1},\mathbf{1})\oplus (\mathbf{1},\mathbf{28},\mathbf{%
1})\oplus (\mathbf{8}_{v},\mathbf{8}_{v},\mathbf{1})\oplus (\mathbf{1},%
\mathbf{1},\mathbf{120})\oplus (\mathbf{8}_{v},\mathbf{1},\mathbf{16})\oplus
(\mathbf{1},\mathbf{8}_{v},\mathbf{16}), \\
\mathbf{32,768} &=&(\mathbf{128},\mathbf{128}^{\prime })\oplus (\mathbf{128}%
^{\prime },\mathbf{128})  \nonumber \\
&=&(\mathbf{8}_{s},\mathbf{8}_{c},\mathbf{128}^{\prime })\oplus (\mathbf{8}%
_{c},\mathbf{8}_{s},\mathbf{128}^{\prime })\oplus (\mathbf{8}_{c},\mathbf{8}%
_{s},\mathbf{128})\oplus (\mathbf{8}_{s},\mathbf{8}_{c},\mathbf{128}).
\label{11-brane}
\end{eqnarray}
\begin{paracol}{2}
\switchcolumn

\begin{figure}
\includegraphics[scale=.5]{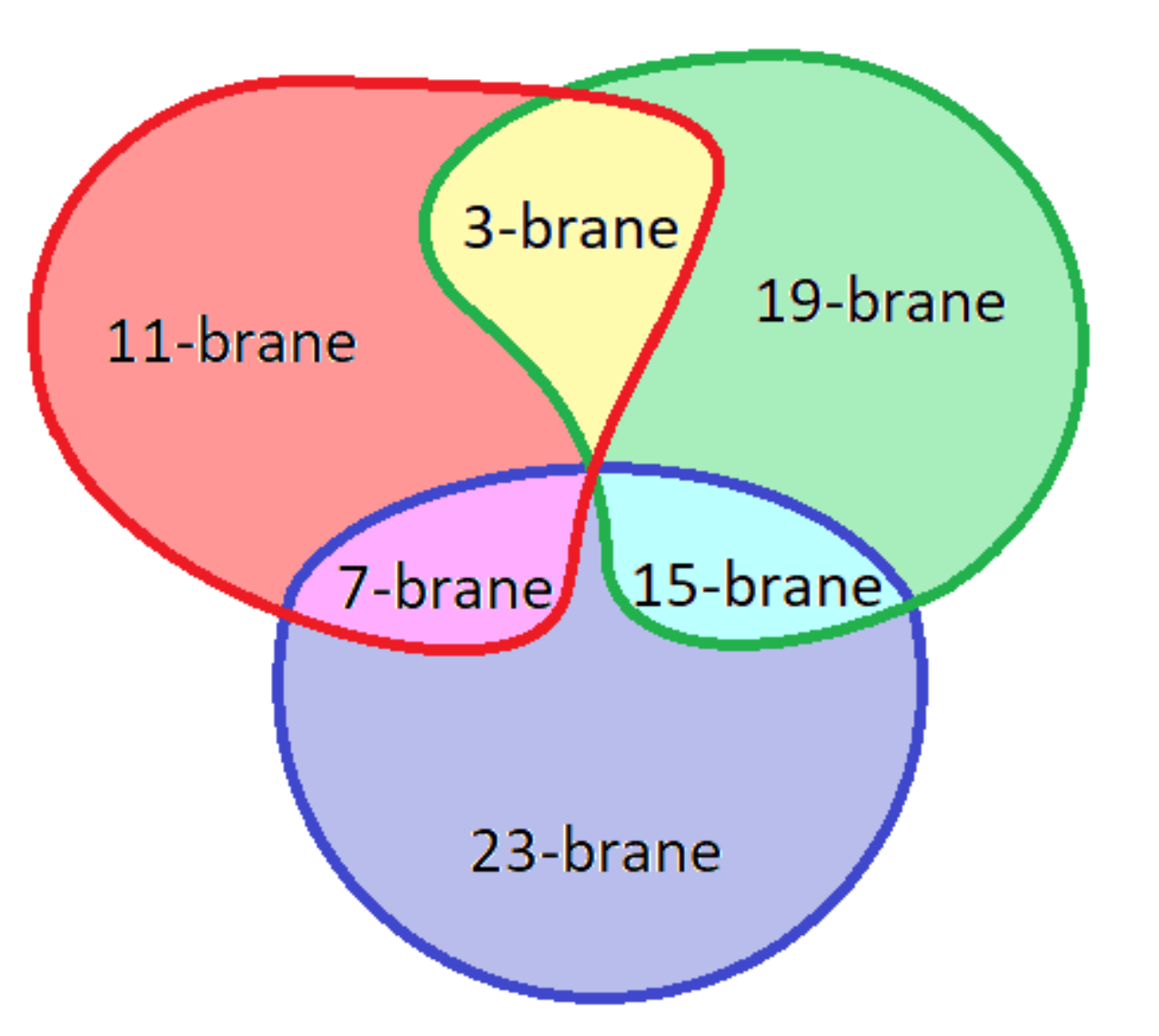}
\caption{The intersections of branes in $D=27+3$. For instance, the 3-brane is dual to the 23-brane, which contains the 7-brane and 15-brane.}
\label{brane-intersections}
\end{figure}
Once the 3-brane worldvolume is isolated within the 11-brane worldvolume, $%
Spin(8)$ is found as being transverse to $Spin(4,4)$ as the conformal
worldvolume symmetry. The 3-brane within the 11-brane within $D=27+3$ has
the effect of making the $Spin(16)$ transverse to the 11-brane hidden with
respect to the matter isolated to the 3-brane and dual 7-brane within $D=11+3
$.

On the other hand, by restricting to the 19-brane worldvolume, the 3-brane
is dual to a 15-brane. Isolating the conformal group of the 19-brane
worldvolume within $\mathfrak{e}_{8(-24)}^{(2)}$ leads to a transverse $%
Spin(8)$,
\vspace{-6pt} 
\end{paracol}
\nointerlineskip
\begin{eqnarray}
\underset{\mathbf{496\oplus 32,768}}{\mathfrak{e}_{8(-24)}^{(2)}}
&\rightarrow &\mathfrak{so}_{28,4}\rightarrow \mathfrak{so}_{20,4}\oplus
\mathfrak{so}_{8}\rightarrow \mathfrak{so}_{4,4}\oplus \mathfrak{so}%
_{16}\oplus \mathfrak{so}_{8}, \\
\mathbf{496} &=&(\mathbf{276},\mathbf{1})\oplus (\mathbf{1},\mathbf{28}%
)\oplus (\mathbf{24},\mathbf{8}_{v})  \nonumber \\
&=&(\mathbf{28},\mathbf{1},\mathbf{1})\oplus (\mathbf{1},\mathbf{120},%
\mathbf{1})\oplus (\mathbf{8}_{v},\mathbf{16},\mathbf{1})\oplus (\mathbf{1},%
\mathbf{1},\mathbf{28})\oplus (\mathbf{8}_{v},\mathbf{1},\mathbf{8}%
_{v})\oplus (\mathbf{1},\mathbf{16},\mathbf{8}_{v}), \\
\mathbf{32,768} &=&(\mathbf{2048},~\mathbf{8}_{c})\oplus (\mathbf{2048}%
^{\prime },~\mathbf{8}_{s})  \nonumber \\
&=&(\mathbf{8}_{s},\mathbf{128}^{\prime },\mathbf{8}_{c})\oplus (\mathbf{8}%
_{c},\mathbf{128},\mathbf{8}_{c})\oplus (\mathbf{8}_{c},\mathbf{128},\mathbf{%
8}_{s})\oplus (\mathbf{8}_{s},\mathbf{128}^{\prime },\mathbf{8}_{s}).
\label{19-brane}
\end{eqnarray}
\begin{paracol}{2}
\switchcolumn

Once the 3-brane worldvolume is isolated within the 19-brane worldvolume, $%
Spin(16)$ is transverse to $Spin(4,4)$. The 3-brane within the 19-brane
in a bulk $D=27+3$ leaves the $Spin(8)$ transverse to the 19-brane hidden with
respect to matter isolated to the 3-brane and its dual 15-brane within $D=19+3$.

While isolating the 3-brane in the 19-brane versus 11-brane symmetries within $%
\mathfrak{e}_{8(-24)}^{(2)}$ leads to different representations, not all
representations of the fermions are identifiable. For instance, the 11-brane
cannot access a 15-brane in its transverse space, such that the $\mathfrak{so}_{16}$
representations are unobservable, while the 19-brane cannot access a 7-brane in its transverse space, such that the $\mathfrak{so}_8$ representations are unobservable.  We assume that the $\mathfrak{so}_8$ transverse to the 19-brane  contains the electromagnetic $\mathfrak{u}_1$ of the standard model~\cite{Slansky:1981yr}.
With the 11-brane and 19-brane intersection, the 3-brane effectively sees
the following fermionic representations with respect to $Spin(4,4)\times
Spin(8)\times Spin(16)$:

\begin{equation}
(\mathbf{8}_s,\mathbf{8}_c,\mathbf{1})\oplus(\mathbf{8}_c,\mathbf{8}_s,%
\mathbf{1}) \oplus (\mathbf{8}_s,\mathbf{1},\mathbf{128}^{\prime }) \oplus (%
\mathbf{8}_c,\mathbf{1},\mathbf{128}),
\end{equation}

\noindent where the singlets effectively arise at low energies, since a
3-brane in an 11-brane cannot probe a transverse 15-brane immersed in $%
D=27+3$, for instance. As such, since the 3-brane is the intersection of two worldvolumes, 128 visible and 2048 dark fermions are found
effectively at low energy from an intersecting braneworld model that contains worldvolumes
that support the commonly studied supergravities and string theories.  The 2048 dark fermions come from $(\mathbf{8}_s,\mathbf{1},\mathbf{128}^{\prime }) \oplus (%
\mathbf{8}_c,\mathbf{1},\mathbf{128})$, which cannot be probed by the $\mathfrak{so}_8$ containing the electromagnetic $\mathfrak{u}_1$.

This number of degrees of freedom corresponds to

\begin{equation}  \label{eq:omegaX}
{\frac{m_{X}}{\mathrm{keV}}}\approx {17.8 {\Omega_{X}h^2} \frac{1.5}{g_{X}}}
,
\end{equation}

\noindent a $\sim$2 keV particle that is in agreement with observations at
both large scales---galaxy formation and distribution---and small scales---cores~\cite{Destri13, Destri13b}.

\section{Conclusions \label{5}}

In this paper, we discuss possible large gauge group representations that allow us to obtain a high number of fermionic degrees of freedom using superalgebras in the context of exceptional periodicity models and M-theory braneworlds. We find that an 11-brane and 19-brane in $D=27+3$ intersect by a 3-brane for space--time such that respective dual 7-branes and 15-branes provide both the degrees of freedom for the standard model and the ones for fermionic WDM, while preserving the ‘separation’ and the lack of interaction between SM and WDM particles. Since $\mathfrak{so_{12,4}}$ combines conformal symmetry with $\mathfrak{so}_{10}$, the $\mathfrak{so}_8$ contains $\mathfrak{su}_3\oplus \mathfrak{u}_1$ of the strong and electromagnetic forces, such that the 19-brane does not interact with light; therefore, it is dark.

In this realization, the intersecting braneworlds span $D=27+3$, but individually span $D=11+3$ and $D=19+3$. For $D=11+3$, the 3-brane is dual to a 7-brane for light matter in 128 off-shell degrees of freedom. For $D=19+3$, the 3-brane is dual to a 15-brane for dark matter. We find 2048 fermionic degrees of freedom in this representation, which correspond to $\sim$2 keV in the model cited in~\cite{bode01}. This particle spectrum is in agreement with constraints from structure formation. Moreover, a 2 keV particle  influences the formation of cores in galaxies~\cite{Destri13, Destri13b}.
Several other representations have been discussed and analyzed. However, this representation has the advantage of naturally allowing for a WDM candidate 
that is interesting from the point of view of astrophysical constraints and also implying worldvolumes that support supergravities and string theories.
The evolution of a particle produced in this manner can be discussed further. While beyond the scope of this paper, further investigations of what happens to these species once the particle decouples would provide additional constraints.

%%%%%%%%%%%%%%%%%%%%%%%%%%%%%%%%%%%%%%%%%%
\vspace{6pt}

%%%%%%%%%%%%%%%%%%%%%%%%%%%%%%%%%%%%%%%%%%
\authorcontributions{S.P. and M.R. initiated this research; S.P. contributed the WDM insights and discussion, M.R. the superalgebras and brane reduction; A.M. contributed to the superalgebras and symmetry breaking; D.C. contributed the braneworld intersection and matter-multiplet phenomenology. All authors contributed to the writing of the paper. All authors have read and agreed to the published version of the manuscript.}

\funding{This research study received no external funding.}

\institutionalreview{Not applicable.}

\informedconsent{Not applicable.}

\acknowledgments{S.P. would like to thank N.G. Sanchez for the invitation to submit to this special issue and for useful suggestions.}

\conflictsofinterest{The authors declare no conflict of interest.}

\abbreviations{Abbreviations}{
The following abbreviations are used in this manuscript:\\

\noindent
\begin{tabular}{@{}ll}
MDPI & Multidisciplinary Digital Publishing Institute\\
DOAJ & directory of open access journals\\
WDM & warm dark matter\\
CDM & cold dark matter\\
SM & standard model\\
EP & exceptional periodicity\\
ESYM & exceptional super Yang--Mills\\
WV & worldvolume
\end{tabular}}

%%%%%%%%%%%%%%%%%%%%%%%%%%%%%%%%%%%%%%%%%%
\end{paracol}
%%%%%%%%%%%%%%%%%%%%%%%%%%%%%%%%%%%%%%%%%%

%%%%%%%%%%%%%%%%%%%%%%%%%%%%%%%%%%%%%%%%%%
\reftitle{References}

%\externalbibliography{yes}
%\bibliography{your_external_BibTeX_file}

%=====================================
% References, variant B: internal bibliography
%=====================================

\end{document}